\documentclass[11pt]{article}

\usepackage[letterpaper,margin=1in]{geometry}
\usepackage{times}
\usepackage{microtype}
\usepackage{amsmath,amssymb,amsfonts}
\usepackage{graphicx}
\usepackage{booktabs}
\usepackage{array}
\usepackage{multirow}
\usepackage{tabularx}
\usepackage{caption}
\usepackage{subcaption}
\usepackage{xcolor}
\usepackage{enumitem}
\usepackage[hidelinks]{hyperref}

\setlist[itemize]{leftmargin=1.3em,itemsep=0.25em,topsep=0.25em}
\newcommand{\ocoot}{OCOO-T }
\newcommand{\pdcorr}{PDCorr}

\title{
  \vspace{-1.0em}\textbf{{\ocoot: A Simple and Scalable Virtual Cell Model for Transcriptional Perturbation Response Prediction}}
}

\author{
\begin{tabular}{c}
Danning Jiang$^{1}$ \quad Zhiwen Yan$^{1}$ \quad Qirun Wang$^{1}$ \quad Zheming An$^{1}$ \\ 
\quad Yalong Zhao$^{1,*}$ \quad Lipeng Lai$^{1,2,*}$\thanks{Corresponding authors: Yalong Zhao and Lipeng Lai. On behalf of the Infevo AI team.}\\
{\small $^{1}$INFevo \quad $^{2}$XtalPi} \\
{\small\texttt{danning.jiang@infevo.ai} \quad \texttt{zhiwen.yan@infevo.ai} \quad \texttt{qirun.wang@infevo.ai} }\\
{\quad \texttt{zheming.an@infevo.ai} \small\texttt{yalong.zhao@infevo.ai}} \\
{ \quad \texttt{lipeng.lai@infevo.ai}, \texttt{lipeng.lai@xtalpi.com}}
\end{tabular}}

\date{}

\begin{document}
\maketitle

\begin{abstract}
  Predicting single-cell transcriptional responses to genetic, chemical and cytokine perturbations is a fundamental challenge in computational biology and AI Virtual Cell (AIVC) modeling, with direct implications for drug discovery and the elucidation of gene regulatory networks. Existing approaches often rely on auxiliary cell-state encoders, hierarchical variational autoencoders, dedicated Transformer encoder-decoder modules, or gene-interaction priors to compress high-dimensional expression profiles into latent representations. While effective, these designs increase architectural complexity and may limit scalability and generalizability. This paper introduces \ocoot\footnote{OCOO is a model series developed by Infinite Evolution Technology Co., Ltd.}, a minimalist flow-matching-based AIVC model for transcriptional perturbation response prediction. OCOO-T utilizes a vanilla Transformer stack that operates directly on continuous gene expression profiles and formulates perturbation response prediction as a continuous-time denoising process. Perturbation embeddings, dosage information, and cell-line/cell-type specificity are integrated through adaptive layer normalization and in-context tokens. Comprehensive evaluations on Tahoe100M, Replogle, and PBMC benchmarks demonstrate that OCOO-T achieves state-of-the-art performance across diverse perturbations and cell types while effectively scaling to long transcriptional profiles through patching and depatching of cellular contexts. By leveraging the simplicity of Transformer-based denoising for single-cell omics, OCOO-T provides an effective and scalable framework for in-silico cellular simulation.
\end{abstract}

\section{Introduction}

Transcriptional perturbation response prediction aims to infer how a cell's gene-expression state changes after external interventions such as gene knockdown, gene knockout, overexpression, chemical treatment, or cytokine stimulation. With the rapid development of single-cell RNA sequencing and pooled perturbation screening technologies, large-scale datasets now measure cellular responses across thousands of genes, diverse perturbation types, multiple dosages, and heterogeneous cellular contexts. These data provide an unprecedented opportunity to model how molecular interventions reshape cellular programs, but experimentally profiling every perturbation-context combination remains infeasible. Accurate computational prediction of perturbation responses is therefore critical for prioritizing candidate interventions, understanding gene regulatory mechanisms, and enabling scalable in-silico cellular simulation.

Earlier efforts to model transcriptional perturbation responses drew on the success of pre-trained language models. scGPT~\cite{cui2024scgpt} pre-trains a generative Transformer on over 33 million human cells with a novel autoregressive gene-masking scheme, demonstrating strong transfer learning across diverse single-cell tasks. However, because single-cell RNA sequencing is destructive, true cell-level correspondences before and after perturbation are fundamentally unobservable. Perturbation response prediction is therefore an inherently distributional problem, requiring models to capture population-level shifts rather than deterministic cell-to-cell mappings. Point-wise prediction objectives, as employed by scGPT and related foundation models, are ill-suited to this setting. STATE~\cite{adduri2025state} directly addressed this challenge by introducing a set Transformer that operates on batches of cells and is trained with a Maximum Mean Discrepancy (MMD) objective, aligning the predicted and observed perturbed cell distributions at the population level rather than minimizing per-cell reconstruction error.

More recently, diffusion- and flow-matching-based approaches~\cite{ho2020ddpm,lipman2023flow} have emerged to generate higher-fidelity transcriptomic profiles. Squidiff~\cite{he2025squidiff} employs a VAE--DDIM hybrid that encodes cells into a semantic latent space and generates perturbed transcriptomes through conditional diffusion decoding, while CellFlow~\cite{klein2025cellflow} and scPPDM~\cite{liang2025scppdm} use conditional flow matching or diffusion model to learn continuous transitions between cellular states in pre-learned latent spaces. Despite their generative capabilities, these methods still operate in VAE-derived latent spaces rather than directly on the original continuous gene-expression profiles. Moreover, for prediction models based on diffusion or flow matching, learning the control-to-perturbation distribution remains a challenge, and distribution-matching strategies based on MMD (e.g., scDFM~\cite{yu2026scdfm}) or optimal transport (e.g., CellFlow) have been extensively explored.

To better handle the absence of cell-level pre/post-perturbation correspondences, PerturbDiff~\cite{yuan2026perturbdiff} models the perturbed cell population as a distribution-valued random variable, defining a diffusion process over kernel mean embeddings of cell populations conditioned on control distributions. While this yields consistent gains across multiple benchmarks and perturbation modalities, the approach faces a fundamental statistical barrier when scaling to long transcriptomic profiles, as the kernel matrix tends to degenerate and the resulting RKHS embeddings of distinct cell populations can become indistinguishable~\cite{ramdas2015decreasing}.

In this work, we propose \ocoot, a minimalist Transformer denoising architecture within the conditional flow matching framework for transcriptional perturbation response prediction. Instead of relying on auxiliary autoencoders or specialized latent spaces, OCOO-T directly denoises continuous gene-expression profile(genetic perturbation), and PBMC~\cite{parsebiosciences2023pbmc} (cytokine stimulation) datasets, OCOO-T demonstras with a plain Transformer backbone conditioned on perturbation and cellular covariates. Across Tahoe100M~\cite{zhang2025tahoe100m} (chemical perturbation), Replogle-Nadig~\cite{nadig2025transcriptomewide} tes strong perturbation response prediction performance across diverse perturbation modalities. A key advantage of OCOO-T is that it supports long expression profiles through a simple patch/unpatch mechanism: the gene-token sequence is patched before Transformer processing and unpatched back to the original transcriptomic resolution afterward. This minimal design enables efficient modeling of long transcriptomic inputs without introducing specialized long-sequence modules or additional auxiliary architectures. Together with this architectural simplicity, our empirical results show that direct covariate conditioning in the denoising network is sufficient to steer generation toward the appropriate target distribution. This further supports the central design principle of OCOO-T: effective perturbation response modeling can be achieved without complex auxiliary distribution-alignment modules.

\enlargethispage{\baselineskip}
\section{Related Work}

\subsection{Deep Learning for Single-Cell Perturbation Prediction}
Predicting the transcriptional responses of single cells to genetic and chemical perturbations is a fundamental challenge in computational biology. Early computational approaches heavily relied on linear models or shallow architectures, which struggled to capture the highly non-linear and context-dependent gene regulatory dynamics. The advent of deep generative models led to a paradigm shift. scGen~\cite{lotfollahi2019scgen} first leveraged variational autoencoders (VAEs) to predict single-cell gene expression under perturbations by performing latent space vector arithmetic. Building on this, the Compositional Perturbation Autoencoder (CPA)~\cite{lotfollahi2023cpa} and its drug-focused variant chemCPA~\cite{hetzel2022chemcpa} introduced adversarial training to disentangle basal cell states, perturbation effects, and covariates (e.g., cell type, dosage, time), enabling the prediction of combinatorial drug effects. Concurrently, GEARS~\cite{roohani2024gears} incorporated biological priors, such as gene-gene co-expression networks and Gene Ontology, to improve generalization to unseen multi-gene perturbations.

More recently, the “pre-train universally, fine-tune on demand” paradigm has emerged with singlecell foundation models. scGPT~\cite{cui2024scgpt} pre-trained a generative Transformer on over 33 million cells, introducing a novel autoregressive attention masking scheme that treats genes as tokens to predict masked expression values. Similarly, scFoundation~\cite{hao2024scfoundation} scaled to 100 million parameters to model large-scale cell atlases. While these foundation models excel at learning generalizable cellular representations, point-wise reconstruction losses in these models fail to capture population-level distributional shifts. To address this, STATE~\cite{adduri2025state} introduced a multi-scale set Transformer that operates on populations of cells rather than individual cells. By training with a Maximum Mean Discrepancy (MMD) objective, STATE aligns the predicted and observed perturbed cell distributions at the population level, enabling robust cross-context generalization.

Building on the set-level modeling paradigm pioneered by STATE, X-Cell~\cite{wang2026xcell} further advances population-level perturbation prediction by combining set-based cell encoding with a diffusion language model framework. Like STATE, X-Cell operates on sets of cells and bypasses the cell-pairing problem entirely through distribution-matching objectives. The model is trained with a composite loss of seven terms targeting distinct statistical properties of the predicted cell distribution, with MMD serving as the core distributional alignment signal. Rather than a single-pass prediction, X-Cell iteratively refines control-to-perturbed state transitions through a masking-based discrete diffusion process: at each training step, a random fraction of control gene expression values are replaced with ground-truth perturbed values, and the model learns to progressively denoise the revealed expression profiles. Gene representations are further enriched with multi-modal biological priors injected via cross-attention. Trained on X-Atlas/Pisces (25.6 million perturbed cells across 16 cellular contexts), X-Cell scales to 4.9 billion parameters and demonstrates power-law scaling behavior, as well as zero-shot generalization to unseen cell types via test-time adaptation on unlabeled control cells.

\subsection{Diffusion and Flow Matching for Biological Generation}
Since 2025, generative diffusion models and conditional flow matching~\cite{ho2020ddpm,lipman2023flow} have rapidly become a major research focus for biological generation, with a growing body of studies showing that these continuous-time generative frameworks can produce higher-fidelity and more distributionally faithful biological profiles than earlier approaches. These continuous-time generative models are uniquely suited to modeling the complex, stochastic trajectories of cellular transitions under perturbations. CellFlow~\cite{klein2025cellflow} utilizes conditional flow matching to map continuous cellular state transitions through time and space. Squidiff~\cite{he2025squidiff} integrates a VAE encoder with a conditional DDIM decoder, compressing cellular expression profiles into a semantic latent variable and a stochastic subcode to generate perturbed transcriptomes via conditional diffusion. Similarly, scPPDM~\cite{liang2025scppdm} performs continuous-time diffusion entirely within a VAE-derived latent space augmented with temporal embeddings.

To more directly address the distributional fitting problem from unpaired data, PerturbDiff~\cite{yuan2026perturbdiff} treats the perturbed cell population itself as a distribution-valued random variable. By embedding empirical cell populations as points in a reproducing kernel Hilbert space (RKHS) via kernel mean embeddings, PerturbDiff defines a DDPM-style diffusion process directly over these RKHS embeddings. This allows the model to capture a manifold of plausible response distributions induced by unobserved latent factors (e.g., microenvironmental fluctuations and batch effects) rather than collapsing to a single expected response.

However, a recurring limitation of many existing diffusion-based models is their reliance on a two-stage pipeline. To bypass the high dimensionality of transcriptomic data, these methods first train a VAE or autoencoder to compress expression profiles and then define the diffusion or flow process in the resulting low-dimensional latent space. Although this strategy helps mitigate the curse of dimensionality, it increases architectural complexity, introduces potential information bottlenecks, and prevents end-to-end optimization in the original expression space. Although PerturbDiff does not rely on a pre-trained autoencoder, it faces a different statistical barrier when scaling to long transcriptomic profiles. Its core training signal and generative representation are both based on kernel mean embeddings in an RKHS, where each cell population is represented by an empirical kernel mean over its member cells. As the dimensionality of cell vectors grows toward the full transcriptome, kernel-based distributional representations become increasingly unreliable. Reddi et al.~\cite{ramdas2015decreasing} formally establish that the statistical power of kernel and distance-based two-sample tests, including those using energy distance kernels, decreases polynomially with dimension under fair alternatives, and that standard bandwidth selection heuristics break down in high dimensions. Since PerturbDiff's diffusion space and training signal are both grounded in such kernel evaluations, this high-dimensional degradation directly corrupts the representational quality of the learned distribution manifold.

\subsection{Long Transcriptomic Profile Generation}
More than 30,000 genes are annotated in the human genome, and many remain poorly characterized \cite{mudge2025gencode}. This motivates long-gene models that can capture transcriptional features across the genome. However, the vast majority of existing single-cell perturbation models restrict their input and output spaces to a predefined set of highly variable genes (HVGs), typically ranging from 2,000 to 5,000 genes. This restriction is primarily driven by the quadratic computational complexity of standard self-attention mechanisms, as well as the optimization challenges associated with fitting high-dimensional, sparse, and zero-inflated distributions. However, restricting the output to a small subset of HVGs severely limits the biological utility of in-silico simulations, as many critical downstream pathways, transcription factors, and low-abundance markers are excluded from consideration.

Recent work has made progress in extending single-cell foundation models to full-transcriptome resolution. scFoundation~\cite{hao2024scfoundation} was among the first to tackle this challenge. Rather than modifying the attention mechanism itself, it exploits the inherent sparsity of scRNA-seq data through an asymmetric encoder--decoder architecture: the encoder processes only non-zero, non-masked gene embeddings (typically a few hundred to a few thousand tokens per cell), while the decoder reconstructs all 19,264 genes. This design concentrates computation on information-rich expressed genes without altering the underlying \(O(N^2)\) attention complexity. However, this sparsity-exploiting design introduces systematic information loss: by excluding zero-expressed and low-abundance genes from the encoder's attention computation, scFoundation's cell-state representations are constructed solely from highly expressed genes. Low-expression genes, including many transcription factors, signaling receptors, and cell-type-specific markers that are expressed at low but functionally critical levels, are entirely absent from the encoder's contextual modeling.

scLong~\cite{bai2026sclong} provides a more direct algorithmic solution to the long-sequence problem. To make it feasible to compute self-attention across all 28,000 genes in the human genome, scLong replaces standard softmax attention with Performer, a linear-complexity attention mechanism based on random feature approximations of the softmax kernel, reducing the attention cost from \(O(N^2)\) to \(O(N)\). Furthermore, scLong introduces a hierarchical dual-encoder design: high-expression genes---which carry core biological information---are processed by a larger Performer encoder with more layers and parameters, while low-expression genes are handled by a lightweight mini-Performer encoder; the outputs of both are then fused by a final full-length Performer for genome-wide contextualization. However, the use of Performer linear attention introduces a fundamental approximation trade-off: by replacing the exact softmax kernel with a random feature map, Performer sacrifices the precise pairwise attention scores that are critical for capturing sharp, localized gene--gene interactions.

\section{Method}

\subsection{Problem Setup}
OCOO-T is a transcriptomic response prediction model designed to predict cellular responses across heterogeneous perturbation modalities. Let $\mathcal{G} = \{g_1, g_2, \ldots, g_G\}$ denote the set of G profiled genes. The control (pre-perturbation) cell state is represented as $x_{\mathrm{c}} = (x_{\mathrm{c}}(g_1), \ldots, x_{\mathrm{c}}(g_G)) \in \mathbb{R}_{+}^{G}$, where $x_{\mathrm{c}}(g_k)$ is expression value of $g_k$ in the control cell. Similarly, the perturbed cell state is represented as $x_{\mathrm{p}} = (x_{\mathrm{p}}(g_1), \ldots, x_{\mathrm{p}}(g_G)) \in \mathbb{R}_{+}^{G}$, where $x_{\mathrm{p}}(g_k)$ is expression value of $g_k$ in the perturbed cell state. $p \in \{1,...,P\}$ denotes perturbation, and $c \in \{1,...,C\}$, $b \in \{1,...,B\}$ denote the covariates of cell-line (or cell-type) and batch (such as plate resplicate or donor) labels respectively.

Given the above notations, a data sample of perturbation response prediction model is represented as $(x_{\mathrm{c}}, x_{\mathrm{p}}, p, c, b)$, and the dataset is represented as:

\begin{equation}
  \mathcal{D}=\{ (x_{\mathrm{c}}^{(i)}, x_{\mathrm{p}}^{(i)}, p_i, c_i, b_i)\}_{i=1}^N
\end{equation}

Due to limitations of current single-cell experimental technologies, $x_{\mathrm{c}}$ and $x_{\mathrm{p}}$ can not be observed as true cell-level pairs; instead, they are treated as pseudo-paired samples drawn from matched control and perturbed populations under the same experimental covariates. OCOO-T learns the generative model $f_{\theta}(x|p, c, b)$ by capturing the conditional distribution under the combination of perturbations and covariates. These covariates can be provided either explicitly, through learned embeddings injected into the prediction network, or implicitly, through the associated control-cell population used as cellular contexts.

\subsection{Conditional Flow Matching}
OCOO-T formulates the prediction of perturbation responses as a continuous-time generative modeling problem. Let $x \in \mathbb{R}^{G}$ denote the target gene expression profile of a perturbed cell, where $G$ is the number of selected genes. The generation process is conditioned on the perturbation identity $p$, the cell-line context $l$, and an optional batch factor $b$.

Instead of employing traditional diffusion models governed by stochastic differential equations, OCOO-T adopts the Rectified Flow framework to construct a deterministic probability path. During training, we define a linear interpolation between pure Gaussian noise $\epsilon \sim \mathcal{N}(0, \sigma_{\epsilon}^2 I)$ and the target perturbation response $x$:

\begin{equation}
  z_t = t\cdot x + (1-t)\cdot \epsilon,
\end{equation}

where $t \in [0,1]$ is sampled from a logit-normal distribution, i.e.\ $\operatorname{logit}(t) \sim \mathcal{N}(\mu, \sigma_{t}^2)$, with $\mu = -0.8$ by default. The ground-truth velocity field of this linear trajectory is constant with respect to $t$:
\begin{equation}
  v = x - \epsilon.
\end{equation}

A central design choice in flow-based generative models is the prediction space of the neural network---what the network directly outputs. Following JiT~\cite{li2025back}, we consider three canonical choices:

\begin{itemize}
  \item \textbf{$x$-prediction:} the network directly outputs the perturbation response. The velocity is then derived as $v_{\theta} = (x_{\theta} - z_t)/(1-t)$.
  \item \textbf{$v$-prediction:} the network directly outputs the velocity. The perturbation response is recovered as $x_{\theta} = z_t + (1-t)\cdot v_{\theta}$.
  \item \textbf{$\epsilon$-prediction:} the network directly outputs the noise. The perturbation response is recovered from the predicted noise accordingly.
\end{itemize}

Independently of the prediction space, the training loss can be defined in any of the three spaces ($x$, $v$, or $\epsilon$). Although the three parameterizations are algebraically convertible for exact predictions, the corresponding regression objectives are not equivalent in practice. Changing the loss space induces a timestep-dependent reweighting of the prediction error. Specifically, the $v$-loss and $x$-loss are related by

\begin{equation}
  L_v = \mathbb{E}\lVert v_{\theta} - v \rVert^2 = \mathbb{E}\left[\frac{1}{(1-t)^2}\lVert x_{\theta} - x \rVert^2\right].
\end{equation}

The $v$-loss up-weights the contribution of large-$t$ timesteps (i.e., near-clean samples), which empirically leads to better generation quality. $v$-loss can be combined with either $v$-prediction or $x$-prediction. For the former case (i.e. $v$-prediction):

\begin{equation}
  L_v = \mathbb{E}_{t,x,\epsilon}\, \lVert v_{\theta}(z_t, p, c, b, t) - v \rVert^2,
  \qquad
  v_{\theta} = \operatorname{net}_{\theta}(z_t, p, c, b, t).
\end{equation}

And for the latter case (i.e. $x$-prediction):

\begin{equation}
  L_v = \mathbb{E}_{t,x,\epsilon}\, \lVert v_{\theta}(z_t, p, c, b, t) - v \rVert^2,
  \qquad
  v_{\theta} = \frac{\operatorname{net}_{\theta}(z_t, p, c, b, t) - z_t}{1-t}.
\end{equation}

At inference time, the predicted velocity field is integrated via an ODE solver (Euler or Heun, 50 steps by default) to generate the perturbed expression profile $\hat{x}$ from pure noise.

\subsection{Transformer Denoising Network}

\begin{figure}[ht]
  \centering
  \includegraphics[width=0.92\linewidth]{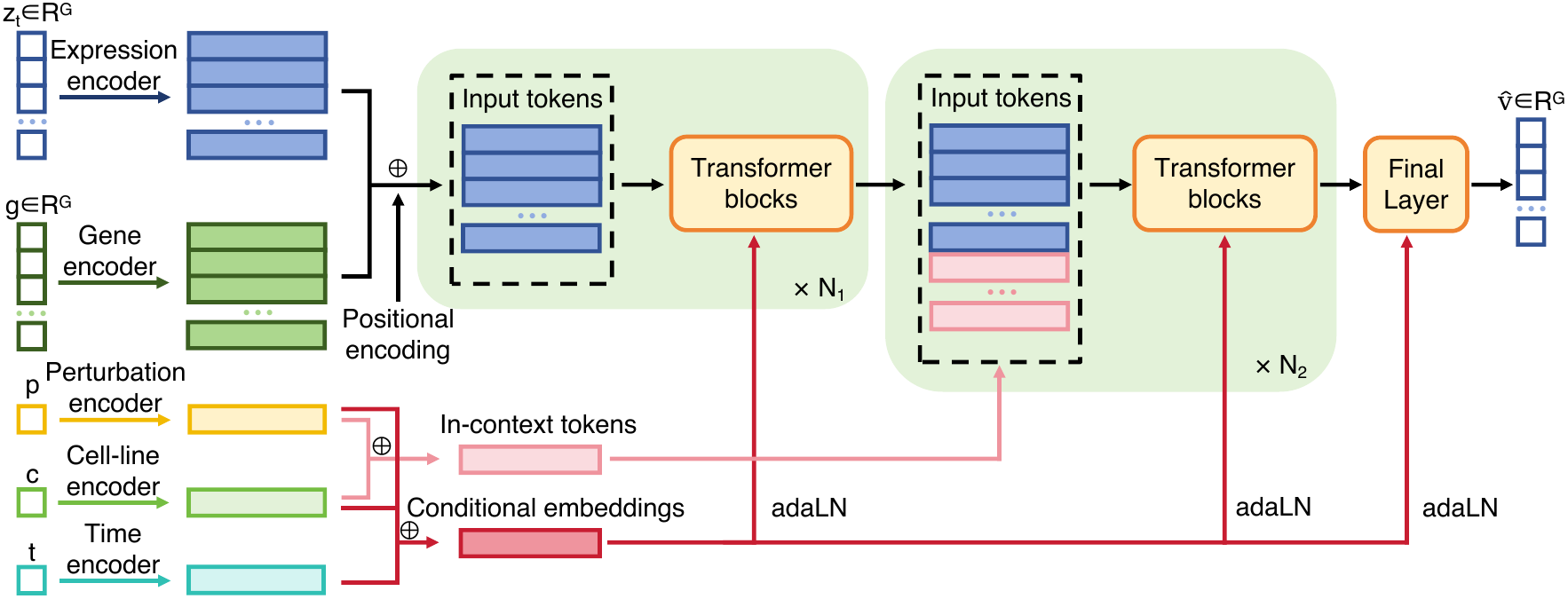}
  \caption{Overview of \ocoot. A continuous expression profile is denoised by Transformer blocks conditioned on perturbation identity and cellular context.}
  \label{fig:overview}
\end{figure}

The core of OCOO-T is a plain Transformer, which is highly scalable and handles high-dimensional gene expression vectors efficiently. Figure 1 shows the network architecture.

A set of general-purpose Transformer improvements that were originally developed for large language models (LLM) are incorporated. This design philosophy ensures that the architecture remains task-agnostic and can benefit from future advances in the broader Transformer literature.

The specific improvements adopted are as follows:

\begin{itemize}
  \item \textbf{SwiGLU FFN.} OCOO-T replaces the standard ReLU feed-forward network with a SwiGLU feed-forward layer. The gated activation $\mathrm{SiLU}(W_1 x) \odot W_2 x$ improves the expressivity of the Transformer block while preserving a simple and scalable architecture.
  \item \textbf{RMSNorm.} OCOO-T uses RMSNorm in normalization positions instead of conventional LayerNorm. This normalization is computationally efficient and improves training stability in deep Transformer stacks.
  \item \textbf{$qk$-norm.} OCOO-T applies RMSNorm independently to the query and key representations before computing attention scores. This stabilizes attention logits and helps prevent attention-score explosion when modeling long gene-token sequences.
\end{itemize}

\textbf{Continuous Value Encoding.} Unlike existing models that rely on discrete binning or rank-based tokenization, OCOO-T directly processes continuous gene expression values. The noisy expression $z_t$ is projected into the model's hidden dimension $D$ using a two-layer multi-layer perceptron (MLP):

\begin{equation}
  h_{\mathrm{expr}} = \mathrm{MLP}(z_t) \in \mathbb{R}^{G \times D}.
\end{equation}

This continuous encoding preserves the numerical precision of the transcriptomic profile without introducing artificial quantization boundaries.

\textbf{Gene Identity Embedding.} To inject biological semantics, each gene is assigned a pre-trained identity embedding. We utilize frozen representations extracted from pretrained large-scale protein language models (e.g., ESM2). These 5120-dimensional embeddings are linearly projected to the hidden dimension $D$ and added to the expression tokens as semantic priors:

\begin{equation}
  h_{\mathrm{gene}} = W_{\mathrm{gene}} \cdot \mathrm{ESM2}(\mathrm{gene\_indices}) \in \mathbb{R}^{G \times D},
\end{equation}

Then each gene token in the transcriptional profile is represented as follows:

\begin{equation}
  x_{\mathrm{in}} = h_{\mathrm{expr}} + h_{\mathrm{gene}},
\end{equation}

The resulting $x_{\mathrm{in}}$ serves as the main token sequence input to the Transformer.

\textbf{In-Context Condition Tokens.} To further strengthen the conditioning, especially for complex dual-condition scenarios (e.g., a specific perturbation in a specific cell line), OCOO-T prepends a sequence of learnable "in-context tokens" to the gene token sequence at a designated intermediate layer. These tokens explicitly carry the perturbation and cell-line embeddings, allowing the self-attention mechanism to route condition-specific information dynamically.

\textbf{Conditioning via Adaptive Layer Normalization (AdaLN).} The conditioning signals---including the timestep $t$, perturbation identity $y$, cell-line $c$, and dosage $d$---are independently embedded and aggregated into a unified condition vector $c_{\mathrm{emb}}$. The perturbation embeddings can be derived from pre-trained molecular fingerprints (e.g., Morgan fingerprints for drugs) or gene embeddings (for genetic knockouts). The aggregated condition vector modulates the Transformer blocks via AdaLN:

\begin{equation}
  \mathrm{AdaLN}(h, c_{\mathrm{emb}}) = x \cdot (1 + \gamma) + \beta,
\end{equation}

where $h$ is Transformer block output to be modulated,  $\gamma$ and $\beta$ are regressed from $c_{\mathrm{emb}}$ using an MLP. This mechanism efficiently broadcasts the global perturbation context to all gene tokens.

\textbf{Dual-Conditional Classifier-Free Guidance.}
To enhance the fidelity of the generated transcriptomes and enforce strict adherence to both the perturbation and cell-line conditions, OCOO-T employs classifier-free guidance (CFG) during inference.

During training, we implement a joint dropout strategy where the perturbation label $y$ and the cell-line label $c$ are independently dropped (replaced by null embeddings) with probability $p_{\mathrm{drop}}$, covering three dropout modes: drop $c$ only, drop $y$ only, or drop both. This trains the model to produce conditional and unconditional velocity estimates simultaneously.

During ODE-based sampling, the predicted velocity is extrapolated using a dual-conditional CFG formula that explicitly balances guidance from both conditions:

\begin{equation}
  \tilde{v} = v_{\mathrm{uncond}} + \frac{w}{2}\left(v_{\mathrm{cond-}c} - v_{\mathrm{uncond}}\right) + \frac{w}{2}\left(v_{\mathrm{cond-}y} - v_{\mathrm{uncond}}\right) + w\left(v_{\mathrm{cond-both}} - v_{\mathrm{uncond}}\right).
\end{equation}

Here $w$ is the guidance scale, $v_{\mathrm{cond-}c}$ and $v_{\mathrm{cond-}y}$ denote the velocity conditioned on cell-line or perturbation alone, respectively, and $v_{\mathrm{cond-both}}$ is the fully conditional estimate. CFG is applied only within a designated timestep interval $[t_{\min}, t_{\max}]$ to avoid over-guidance at extreme noise levels. In addition, two exponential moving average (EMA) copies of the model weights, with decay rates 0.9999 and 0.9996, are maintained throughout training, and the EMA model is used exclusively for inference.

\subsection{Condition Representation and Injection}
\begin{figure}[ht]
  \centering
  \includegraphics[width=0.92\linewidth]{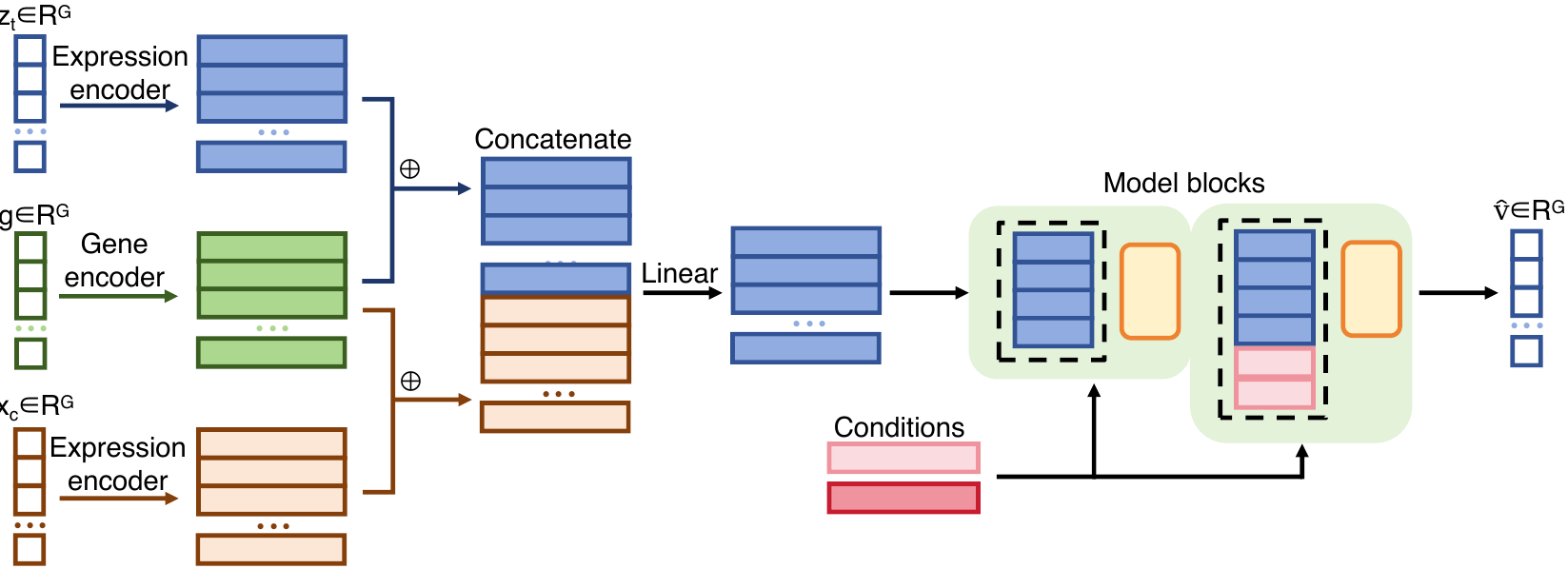}
  \caption{Control cells ($\textbf x_c$) are injected along with the input for perturbation prediction.}
  \label{fig:ctrl2Pert}
\end{figure}

Figure 1 illustrates the default condition-representation setup in OCOO-T, where perturbation identity, cell-line or cell-type identity, and batch information are encoded as label-derived embedding vectors.

For perturbation conditioning, we use modality-specific representations:
\begin{itemize}
  \item Chemical perturbations in the Tahoe100M benchmark are represented by molecular fingerprints. Dosage is encoded with a continuous value encoder, and the resulting dosage vector is added to the molecular representation to capture dose-dependent chemical effects.
  \item Genetic perturbations in the Replogle-Nadig benchmark share the same embedding space as gene tokens; in our experiments, these gene embeddings are initialized from ESM2 representations.
  \item Cytokine stimulations in the PBMC benchmark are represented using ESM2 protein embeddings. For protein complexes, the final representation is obtained by pooling the ESM2 embeddings of all component proteins.
\end{itemize}

In the default setup, cell-line or cell-type identities are represented by learnable embedding parameters. For the PBMC benchmark, donor identities are additionally embedded and incorporated into the overall condition representation.

To better support practical inference scenarios in which the target cellular context may be absent from training, we further introduce an ``implicit'' cellular condition injection strategy. As shown in Figure 2, the control-cell context $x_c$ is provided to the Transformer together with the noisy state $z_t$. Both $x_c$ and $z_t$ are first processed by the Gene Value Encoder:

\begin{equation}
  h_{\mathrm{expr}}(z_t) = \mathrm{MLP}(z_t), \qquad
  h_{\mathrm{expr}}(x_c) = \mathrm{MLP}(x_c).
\end{equation}

Then the value vectors are added to gene tokens:

\begin{equation}
  x_{\mathrm{in}}(z_t) = h_{\mathrm{expr}}(z_t) + h_{\mathrm{gene}}, \qquad
  x_{\mathrm{in}}(x_c) = h_{\mathrm{expr}}(x_c) + h_{\mathrm{gene}}.
\end{equation}

Finally $x_{\mathrm{in}}(z_t)$ and $x_{\mathrm{in}}(x_c)$ are concatenated and projected to the original feature dimension:

\begin{equation}
  x_{\mathrm{proj}} = \mathrm{MLP}_{\mathrm{out}}\!\left(\mathrm{Concat}\bigl(x_{\mathrm{in}}(z_t), x_{\mathrm{in}}(x_c)\bigr)\right).
\end{equation}

Because single-cell perturbation assays do not provide true cell pairs before and after intervention, perturbation response prediction is naturally formulated as a control-to-perturbed distribution transformation rather than a deterministic cell-to-cell mapping. Accordingly, we represent each cellular population using the mean profile of a set of control cells, where a population is defined by a specific combination of perturbation, cell line or cell type, and batch. Under this setting, the cellular context condition no longer corresponds to a single cell; instead, it summarizes the distributional characteristics of a group of cells and provides population-level context for predicting the corresponding perturbed distribution.

\subsection{Scaling to Long Transcriptional Profiles}
\begin{figure}[ht]
  \centering
  \includegraphics[width=0.92\linewidth]{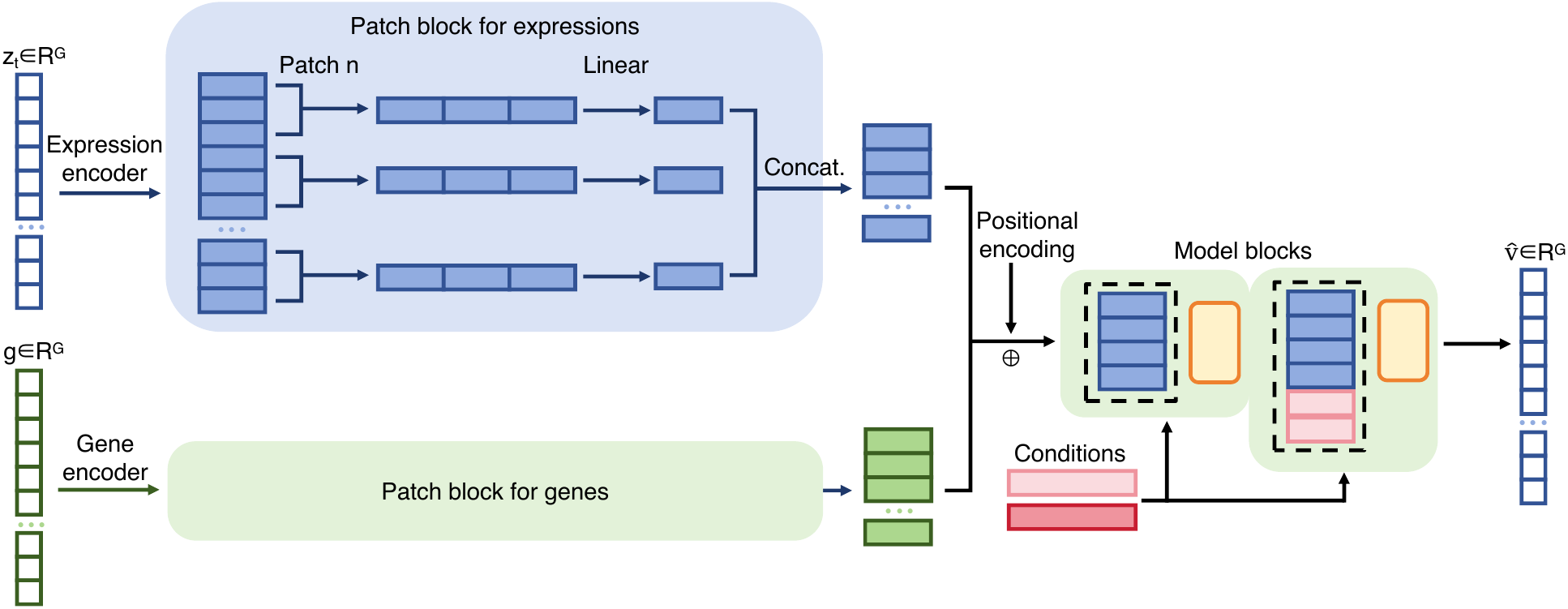}
  \caption{Patching enables the modeling of long-panel genes.}
  \label{fig:patch}
\end{figure}

Directly applying self-attention to full transcriptomic profiles is computationally expensive, because the cost of self-attention grows quadratically with sequence length, and the sequence length itself scales with the number of modeled genes. To make OCOO-T applicable to long gene panels, we adopt a simple patching and depatching strategy along the gene dimension (Figure~\ref{fig:patch}). Specifically, ordered gene-level expression values and gene embeddings are grouped into contiguous patches of size \(s\). Each patch is projected into a single patch token before Transformer processing, and the Transformer outputs are subsequently depatched back to gene-level predictions. This design preserves the plain Transformer backbone while reducing the effective attention length from the number of genes \(G\) to approximately \(G/s\).

The patching operation is applied consistently to the noisy expression state, gene-level representations, and cellular-context inputs when control-cell profiles are used as conditions. After denoising, the depatching module reconstructs predictions at the original gene resolution, so the model can still generate full-panel transcriptional profiles rather than only low-dimensional summaries. We evaluate patch sizes of 8, 16, and 32 on 18,080-gene profiles from AIVC~\cite{roohani2025virtualcellchallenge}; to make the patching/depatching experiments directly comparable with the other experimental setups, we report all benchmark metrics on the same 2,000 HVGs used elsewhere in the study.

\section{Experiment}

\subsection{Benchmark Datasets and Splits}

We use Tahoe100M, Replogle-Nadig, and PBMC as established benchmarks for transcriptional response prediction, which have been widely adopted as standard benchmarks in the literatures. All benchmarks are evaluated on 2,000 highly variable genes (HVGs) except the experiments for long transcriptomic profiles, and their preprocessing and train/validation/test splits follow the STATE protocol to enable fair comparison across methods~\cite{adduri2025state,yuan2026perturbdiff}. Details are listed in Table~\ref{tab:benchmark_summary}.

\begin{table}[htb]
  \centering
  \caption{Updated summary of benchmark dataset sizes (validated from raw h5ad files).}
  \label{tab:benchmark_summary}
  \resizebox{\textwidth}{!}{%
  \begin{tabular}{cccccc}
    \toprule
    \multirow{2}{*}{Benchmark} & \multirow{2}{*}[-0.6ex]{\shortstack[c]{Perturbation\\Modality}} & \multicolumn{3}{c}{Perturbed cells} & \multirow{2}{*}{Control cells} \\
    \cmidrule(lr){3-5}
    & & \multicolumn{1}{c}{Train} & \multicolumn{1}{c}{Validation} & \multicolumn{1}{c}{Test} & \\
    \midrule
    \raisebox{\dimexpr1.4\normalbaselineskip-\height}{\shortstack[c]Tahoe100M} &
    \raisebox{\dimexpr1.4\normalbaselineskip-\height}{\shortstack[l]{Chemical}} &
    \shortstack[c]{89,495,239 cells\\1,137 perturbations\\50 cell-lines} &
    \shortstack[c]{553,076 cells\\54 perturbations\\5 cell-lines} &
    \shortstack[c]{8,270,319 cells\\735 perturbations\\5 cell-lines (same as in validation)} &
    \shortstack[c]{2,330,156 cells\\50 cell-lines} \\
    \midrule
    \raisebox{\dimexpr1.75\normalbaselineskip-\height}{\shortstack[c]{Replogle-\\Nadig}} &
    \raisebox{\dimexpr1.5\normalbaselineskip-\height}{\shortstack[l]{Genetic}} &
    \shortstack[c]{572,545 cells\\2,013 perturbations\\4 cell-lines} &
    \shortstack[c]{4,825 cells\\60 perturbations\\1 cell-line} &
    \shortstack[c]{26,878 cells\\380 perturbations\\1 cell-line (same as in validation)} &
    \shortstack[c]{39,165 cells\\4 cell-lines} \\
    \midrule
    \raisebox{\dimexpr1.8\normalbaselineskip-\height}{\shortstack[c]PBMC} &
    \raisebox{\dimexpr1.8\normalbaselineskip-\height}{\shortstack[l]{Cytokine}} &
    \shortstack[c]{6,657,336 cells\\90 perturbations\\18 cell-types\\12 donors} &
    \shortstack[c]{150,484 cells\\4 perturbations\\18 cell-types\\4 donors} &
    \shortstack[c]{2,260,453 cells\\62 perturbations\\18 cell-types\\4 donors (same as in validation)} &
    \shortstack[c]{629,701 cells\\18 cell-types\\12 donors} \\
    \bottomrule
  \end{tabular}
  }
\end{table}

\textbf{Tahoe100M.} Tahoe100M is a large-scale chemical perturbation atlas containing over 100 million single-cell expression profiles from 50 human cancer cell lines exposed to more than 1,100 small-molecule drug and dosage combinations~\cite{zhang2025tahoe100m}. As summarized in Table~\ref{tab:benchmark_summary}, the benchmark contains 89,495,239 training cells across 1,137 perturbations and 50 cell lines, 553,076 validation cells across 54 perturbations and 5 cell lines, 8,270,319 test cells across 735 perturbations and 5 cell lines, and 2,330,156 control cells from 50 cell lines. Overall, the dataset covers 1,137 perturbation conditions, 50 cell lines, and 14 experimental batches.

\textbf{Replogle-Nadig.} Replogle-Nadig is a large-scale single-cell genetic perturbation benchmark built from pooled CRISPR interference (CRISPRi) screens across four human cell lines: K562, RPE1, Jurkat, and HepG2~\cite{nadig2025transcriptomewide,gilbert2013crispri}. As summarized in Table~\ref{tab:benchmark_summary}, the benchmark contains 572,545 training cells across 2,013 perturbations and 4 cell lines, 4,825 validation cells across 60 perturbations and 1 cell line, 26,878 test cells across 380 perturbations and 1 cell line, and 39,165 control cells from 4 cell lines. The benchmark includes 2,013 perturbations and 56 experimental batches~\cite{yuan2026perturbdiff}.

\textbf{PBMC.} PBMC is a large-scale cytokine stimulation dataset generated by Parse Biosciences~\cite{parsebiosciences2023pbmc}. It contains approximately 10 million human peripheral blood mononuclear cells collected from 12 healthy donors, covering 18 immune cell types, 90 distinct cytokine perturbation conditions, and a PBS control condition. After quality control, 9,697,974 cells were retained: 6,657,336 training cells across 90 perturbations, 18 cell types, and 12 donors; 150,484 validation cells across 4 perturbations, 18 cell types, and 4 donors; 2,260,453 test cells across 62 perturbations, 18 cell types, and 4 donors; and 629,701 control cells across 18 cell types and 12 donors (Table~\ref{tab:benchmark_summary}). Prior single-cell perturbation analyses have shown that strong perturbations can induce rich transcriptional phenotypes across many genes~\cite{norman2019exploring}, and many cytokine treatments in PBMC induce pronounced transcriptional responses, including the upregulation of more than 50 genes. Owing to its scale, perturbation diversity, donor diversity, and strong cytokine-induced expression changes, PBMC provides a challenging benchmark for evaluating whether models can capture immune signaling responses in primary cells.

For preprocessing, per-cell library-size normalization and a log1p transformation are applied for all datasets. Specially for Replogle, perturbations and cells are filtered by on-target knockdown efficacy to retain samples with sufficiently strong repression of the targeted gene, following STATE.

\subsection{Benchmark Results}

\begin{figure}[ht]
  \centering
  \includegraphics[width=0.92\linewidth]{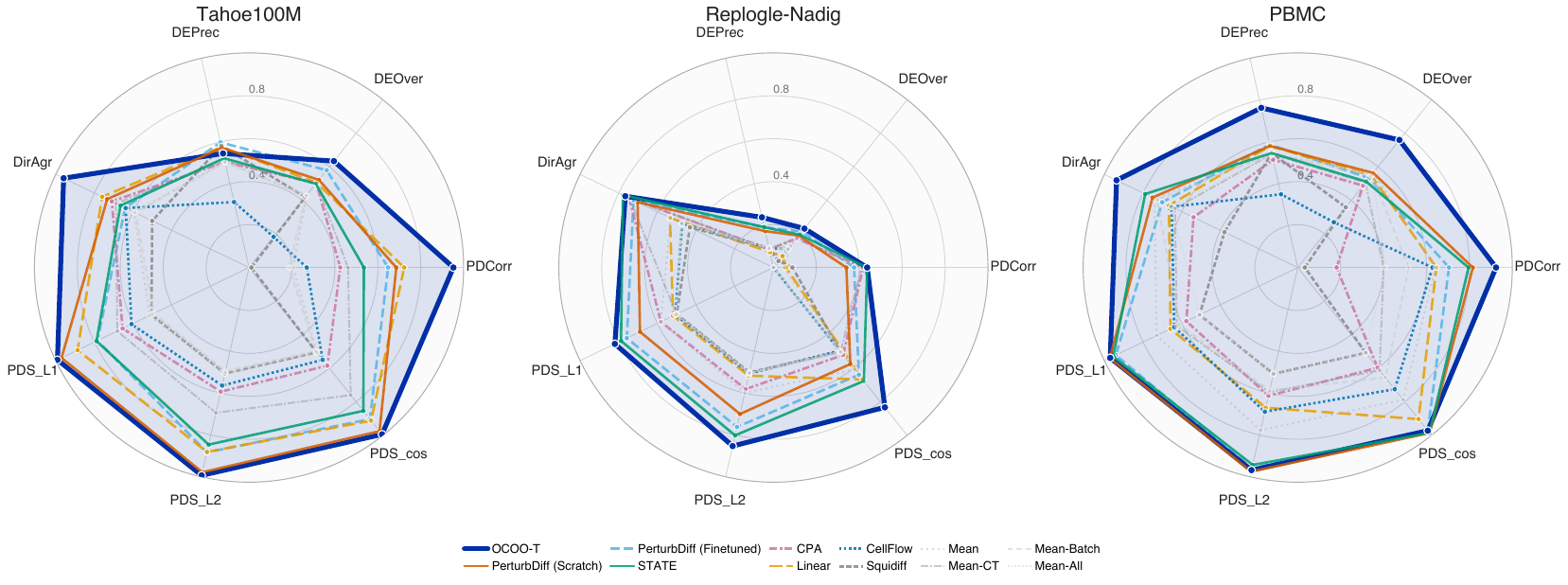}
  \caption{Visualization of multi-dimensional performance across benchmarks.}
  \label{fig:radar}
\end{figure}

Figure~\ref{fig:radar} visualizes the multi-dimensional performance of OCOO-T against baselines across the Tahoe100M, Replogle, and PBMC benchmarks. We evaluate each method using expression-level metrics, including PDCorr, perturbation discrimination scores (PDS), MSE, and MAE, together with differential-expression metrics, including DEOver, DEPrec, and DirAgr. Note that absolute reconstruction errors (MSE and MAE) are omitted from the radar charts due to scaling differences. Detailed quantitative results are deferred to Tables~\ref{tab:tahoe-results}--\ref{tab:pbmc-results}. All metrics are computed via the STATE cell-eval toolkit~\cite{arcinstitute2026celleval}, with detailed definitions and computation procedures provided in Appendix~\ref{app:evaluation_metrics}. We compare OCOO-T against a broad range of strong baseline systems. Most baseline results are taken from~\cite{yuan2026perturbdiff}. We follow the same dataset splits, preprocessing pipelines, and HVG selections to ensure a fair comparison.

We compare OCOO-T with the following systems:
\begin{itemize}
  \item \textbf{PerturbDiff} is a conditional diffusion model that generates the distribution of perturbed single-cell transcriptomes by denoising from a reference control cell population~\cite{yuan2026perturbdiff}. Two variants are presented: PerturbDiff (Scratch) is trained end-to-end using the perturbation data, while PerturbDiff (Finetuned) first performs large-scale marginal pretraining and is then finetuned on perturbation data.
  \item \textbf{STATE} is a population-matching method based on a kernel discrepancy objective~\cite{adduri2025state}.
  \item \textbf{CPA} is a deep generative model for compositional perturbation prediction~\cite{lotfollahi2023cpa}.
  \item \textbf{Linear} is a simple linear baseline for perturbation effect prediction~\cite{ahlmann2024linear}.
  \item \textbf{CellFlow} models perturbation transitions with flow matching~\cite{klein2025cellflow}.
  \item \textbf{Squidiff} is a diffusion-based perturbation predictor~\cite{he2025squidiff}.
  \item \textbf{Mean} predicts by assigning the average observed perturbed profile, and the three mean variants aggregate this average overall, per cell type, or per batch~\cite{kernfeld2025comparison}.
\end{itemize}

The above list includes both state-of-the-art and simple systems. Simple methods such as Linear and Mean sometimes perform surprisingly well, helping indicate what the true baseline should be.

On \textbf{Tahoe100M}, OCOO-T achieves the strongest overall performance across all evaluated methods, obtaining the best PDCorr ($0.952$), reconstruction error (MSE $=1.11\times 10^{-4}$, MAE $=7.27\times 10^{-3}$), all three PDS metrics (PDS\_L1 $=0.991$, PDS\_L2 $=0.993$, PDS\_cos $=0.994$), DEOver ($0.634$), and DirAgr ($0.959$). Within the two PerturbDiff variants, PerturbDiff (Scratch) is stronger on perturbation-discrimination metrics, with PDCorr $=0.686$, PDS\_L1 $=0.970$, PDS\_L2 $=0.978$, and PDS\_cos $=0.975$, whereas PerturbDiff (Finetuned) performs better on DE-oriented metrics, reaching DEOver $=0.580$ and the best DEPrec at $0.598$. The Linear baseline remains competitive on reconstruction error among the non-OCOO-T baselines (MSE $=4.15\times 10^{-4}$, MAE $=0.009$) and attains DirAgr $=0.760$. The Tahoe100M results overall show OCOO-T as the dominant method and the two PerturbDiff variants as trading off discrimination and DE recovery.

On \textbf{Replogle}, OCOO-T again leads most metrics, with the best PDCorr ($0.437$), all three PDS scores (PDS\_L1 $=0.820$, PDS\_L2 $=0.853$, PDS\_cos $=0.833$), and the strongest DE recovery scores (DEOver $=0.232$, DEPrec $=0.240$). STATE is nevertheless competitive: it closely matches OCOO-T on PDCorr ($0.437$ vs. $0.437$), achieves the best MSE ($6.40\times 10^{-3}$), and records the highest DirAgr ($0.778$). Within the PerturbDiff variants, finetuning improves substantially over training from scratch on most selected metrics, especially DEOver ($0.214$ vs. $0.190$), DEPrec ($0.192$ vs. $0.174$), DirAgr ($0.723$ vs. $0.702$), PDCorr ($0.376$ vs. $0.340$), and all three PDS scores. This is consistent with the discussion in PerturbDiff that pretraining is particularly beneficial in more limited-data regimes such as Replogle.

On \textbf{PBMC}, OCOO-T shows particularly strong performance on cytokine-response and DE-oriented metrics, achieving the best PDCorr ($0.924$), DEOver ($0.760$), DEPrec ($0.763$), and DirAgr ($0.937$). OCOO-T also obtains strong PDS metrics that are comparable to the best-performing method. PerturbDiff (Scratch) remains strongest on the three perturbation discrimination scores (PDS\_L1 $=0.972$, PDS\_L2 $=0.975$, PDS\_cos $=0.986$), while STATE achieves the lowest reconstruction error (MSE $=1.41\times 10^{-4}$, MAE $=0.005$). Overall, the PBMC results suggest that OCOO-T is especially effective at recovering cytokine-induced transcriptional directionality and differential-response structure in primary immune cells, while several baselines remain competitive for absolute reconstruction or perturbation-discrimination ranking.

\begin{table}[htb]
  \centering
  \caption{Tahoe100M benchmark results.}
  \label{tab:tahoe-results}
  \resizebox{\textwidth}{!}{%
    \begin{tabular}{lccccccccc}
      \toprule
      Model & \pdcorr $\uparrow$ & DEOver $\uparrow$ & DEPrec $\uparrow$ & DirAgr $\uparrow$  & PDS\_L1 $\uparrow$ & PDS\_L2 $\uparrow$ & PDS\_cos $\uparrow$ & MSE$\downarrow$ & MAE$\downarrow$ \\
      \midrule
      OCOO-T & \textbf{0.952} & \textbf{0.634} & 0.545 & \textbf{0.959} & \textbf{0.991} & \textbf{0.993} & \textbf{0.994} & \textbf{1.11e-4} & \textbf{7.27e-3} \\
      PerturbDiff (Scratch) & 0.686 & 0.522 & 0.572 & 0.734 & 0.970 & 0.978 & 0.975 & 6.30e-4 & 0.012 \\
      PerturbDiff (Finetuned) & 0.648 & 0.580 & \textbf{0.598} & 0.641 & 0.784 & 0.883 & 0.903 & 1.04e-3 & 0.017 \\
      STATE & 0.535 & 0.499 & 0.522 & 0.665 & 0.789 & 0.846 & 0.854 & 2.11e-3 & 0.021 \\
      Mean & 0.205 & 0.430 & 0.504 & 0.595 & 0.515 & 0.508 & 0.508 & 3.09e-3 & 0.026 \\
      CPA & 0.425 & 0.502 & 0.504 & 0.710 & 0.654 & 0.593 & 0.585 & 8.89e-4 & 0.011 \\
      Linear & 0.723 & 0.505 & 0.576 & 0.760 & 0.887 & 0.881 & 0.912 & 4.15e-4 & 0.009 \\
      CellFlow & 0.269 & 0.183 & 0.313 & 0.638 & 0.608 & 0.564 & 0.550 & 3.10e-3 & 0.027 \\
      Squidiff & 0.011 & 0.420 & 0.581 & 0.501 & 0.505 & 0.506 & 0.505 & 5.627 & 2.244 \\
      Mean Variant (per Cell Type) & 0.461 & 0.504 & 0.508 & 0.688 & 0.682 & 0.694 & 0.760 & 6.96e-4 & 0.010 \\
      Mean Variant (per Batch) & 0.185 & 0.427 & 0.504 & 0.587 & 0.492 & 0.496 & 0.498 & 3.11e-3 & 0.026 \\
      Mean Variant (Overall) & 0.193 & 0.429 & 0.504 & 0.591 & 0.501 & 0.501 & 0.501 & 3.06e-3 & 0.026 \\
      \bottomrule
    \end{tabular}%
  }
\end{table}

\begin{table}[htb]
  \centering
  \caption{Replogle-Nadig benchmark results.}
  \label{tab:replogle-results}
  \resizebox{\textwidth}{!}{%
    \begin{tabular}{lccccccccc}
      \toprule
      Model & \pdcorr $\uparrow$ & DEOver $\uparrow$ & DEPrec $\uparrow$ & DirAgr $\uparrow$  & PDS\_L1 $\uparrow$ & PDS\_L2 $\uparrow$ & PDS\_cos $\uparrow$ & MSE $\downarrow$& MAE$\downarrow$ \\
      \midrule
      OCOO-T & \textbf{0.437} & \textbf{0.232} & \textbf{0.240} & 0.766 & \textbf{0.820} & \textbf{0.853} & \textbf{0.833} & 7.90e-3 & 0.061 \\
      PerturbDiff (Scratch) & 0.340 & 0.190 & 0.174 & 0.702 & 0.690 & 0.700 & 0.575 & 1.47e-2 & 0.081 \\
      PerturbDiff (Finetuned) & 0.376 & 0.214 & 0.192 & 0.723 & 0.758 & 0.762 & 0.639 & 1.46e-2 & 0.079 \\
      STATE & \textbf{0.437} & 0.196 & 0.193 & \textbf{0.778} & 0.788 & 0.802 & 0.676 & \textbf{6.40e-3} & 0.055 \\
      Mean & 0.048 & 0.127 & 0.094 & 0.532 & 0.582 & 0.599 & 0.608 & 9.90e-2 & 0.206 \\
      CPA & 0.418 & 0.173 & 0.087 & 0.746 & 0.582 & 0.581 & 0.521 & 7.50e-2 & \textbf{0.054} \\
      Linear & 0.058 & 0.068 & 0.074 & 0.535 & 0.519 & 0.517 & 0.667 & 1.30e-2 & 0.074 \\
      CellFlow & -0.003 & 0.110 & 0.093 & 0.472 & 0.507 & 0.507 & 0.495 & 9.49e-2 & 0.205 \\
      Squidiff & 0.089 & 0.039 & 0.091 & 0.432 & 0.500 & 0.501 & 0.501 & 4.641 & 2.077 \\
      Mean Variant (per Cell Type) & 0.412 & 0.178 & 0.087 & 0.743 & 0.502 & 0.505 & 0.502 & 8.45e-3 & 0.057 \\
      Mean Variant (per Batch) & 0.002 & 0.110 & 0.092 & 0.475 & 0.505 & 0.505 & 0.506 & 6.99e-2 & 0.174 \\
      Mean Variant (Overall) & -0.001 & 0.111 & 0.093 & 0.474 & 0.502 & 0.503 & 0.501 & 7.11e-2 & 0.175 \\
      \bottomrule
    \end{tabular}%
  }
\end{table}

\begin{table}[htb]
  \centering
  \caption{PBMC benchmark results.}
  \label{tab:pbmc-results}
  \resizebox{\textwidth}{!}{%
    \begin{tabular}{lccccccccc}
      \toprule
      Model & \pdcorr $\uparrow$ & DEOver $\uparrow$ & DEPrec $\uparrow$ & DirAgr $\uparrow$  & PDS\_L1 $\uparrow$ & PDS\_L2 $\uparrow$ & PDS\_cos $\uparrow$ & MSE $\downarrow$& MAE $\downarrow$\\
      \midrule
      OCOO-T & \textbf{0.924} & \textbf{0.760} & \textbf{0.763} & \textbf{0.937} & 0.970 & 0.968 & 0.972 & 1.81e-4 & 0.006 \\
      PerturbDiff (Scratch) & 0.816 & 0.564 & 0.581 & 0.751 & \textbf{0.972} & \textbf{0.975} & \textbf{0.986} & 1.91e-4 & 0.006 \\
      PerturbDiff (Finetuned) & 0.705 & 0.533 & 0.588 & 0.701 & 0.941 & 0.959 & 0.972 & 3.15e-4 & 0.008 \\
      STATE & 0.796 & 0.512 & 0.547 & 0.789 & 0.954 & 0.943 & 0.985 & \textbf{1.41e-4} & \textbf{0.005} \\
      Mean & 0.642 & 0.564 & 0.544 & 0.740 & 0.730 & 0.777 & 0.787 & 8.49e-4 & 0.013 \\
      CPA & 0.181 & 0.488 & 0.515 & 0.539 & 0.576 & 0.613 & 0.597 & 1.10e-2 & 0.052 \\
      Linear & 0.646 & 0.549 & 0.581 & 0.666 & 0.658 & 0.670 & 0.903 & 4.44e-4 & 0.009 \\
      CellFlow & 0.628 & 0.270 & 0.350 & 0.652 & 0.639 & 0.689 & 0.725 & 3.58e-4 & 0.009 \\
      Squidiff & 0.033 & 0.359 & 0.547 & 0.379 & 0.508 & 0.508 & 0.508 & 4.387 & 2.062 \\
      Mean Variant (per Cell Type) & 0.400 & 0.506 & 0.542 & 0.635 & 0.625 & 0.596 & 0.612 & 6.39e-4 & 0.009 \\
      Mean Variant (per Batch) & 0.517 & 0.554 & 0.542 & 0.727 & 0.612 & 0.587 & 0.655 & 5.59e-4 & 0.007 \\
      Mean Variant (Overall) & 0.408 & 0.557 & 0.542 & 0.656 & 0.508 & 0.508 & 0.508 & 1.01e-3 & 0.013 \\
      \bottomrule
    \end{tabular}%
  }
\end{table}

\subsection{Long Transcriptomic Profile Generation Results}
We next evaluate whether the proposed patching/depatching strategy enables OCOO-T to scale from the standard 2,000-HVG setting to long transcriptomic profiles on Replogle-Nadig benchmark. Specifically, we train and evaluate OCOO-T on 18,080-gene inputs using patch sizes of 8, 16, and 32, while reporting metrics on the same 2,000 HVGs used in the short-panel benchmark for direct comparability. As shown in Table~\ref{tab:patch}, the patched models retain performance close to the short-panel model across perturbation-effect and differential-expression metrics, despite reducing the Transformer sequence length by a large factor. These results indicate that gene-dimension patching provides an effective computational trade-off: it substantially lowers attention cost while preserving the predictive signal needed for perturbation response modeling.

\begin{figure}[htb]
  \centering
  \includegraphics[width=0.92\linewidth]{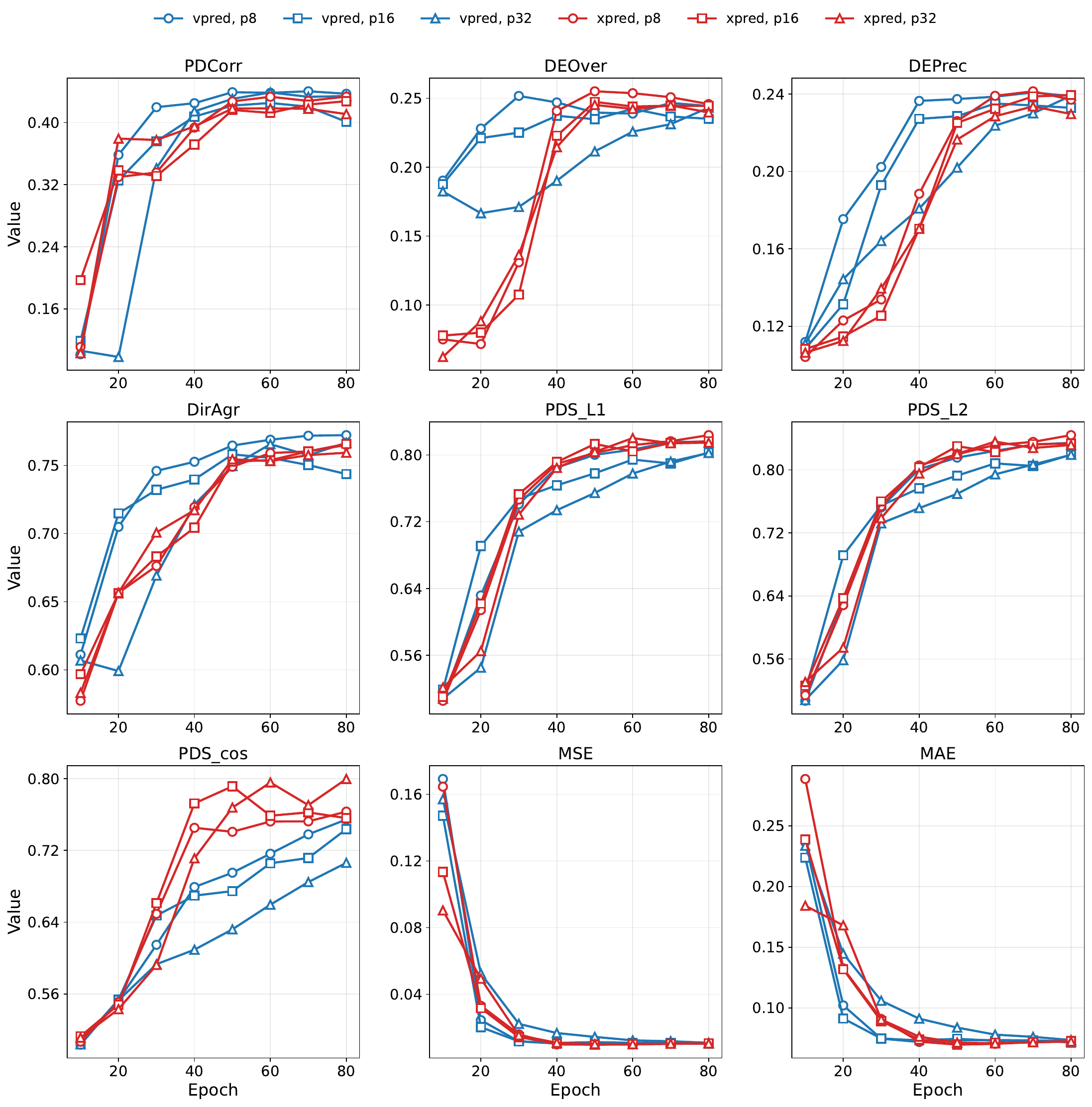}
  \caption{Comparison of the benchmark results between v-prediction and x-prediction under different patch sizes. vpred/xpred, v/x-prediction; p8/16/32, patch size 8/16/32.}
  \label{fig:vx_comp}
\end{figure}

We further compare two denoising parameterizations under the patched setting: velocity prediction (v-prediction), which is standard in flow matching, and direct clean-expression prediction (x-prediction). In the image-generation tasks studied by Li and He~\cite{li2025back}, x-prediction was reported to outperform v-prediction for high-dimensional image patches, potentially because natural images lie on structured low-dimensional manifolds. In OCOO-T, however, x-prediction and v-prediction achieve broadly comparable final performance, although they exhibit different convergence behavior across metrics during training (Figure~\ref{fig:vx_comp}, Table~\ref{tab:patch}). This suggests that the preference for a denoising target is data-domain dependent. Unlike image patches, transcriptomic vectors are sparse, heavy-tailed, biologically heterogeneous, and not organized on a spatially local grid; consequently, the advantage of x-prediction observed for image manifolds does not directly transfer to transcriptomic perturbation modeling.

\begin{table}[htb]
  \centering
  \caption{Replogle-Nadig evaluation results for different patch sizes.}
  \label{tab:patch}
  \small
  \resizebox{\textwidth}{!}{%
    \begin{tabular}{lccccccccc}
      \toprule
      Model & \pdcorr $\uparrow$ & DEOver $\uparrow$ & DEPrec $\uparrow$ & DirAgr $\uparrow$  & PDS\_L1 $\uparrow$ & PDS\_L2 $\uparrow$ & PDS\_cos $\uparrow$ & MSE$\downarrow$ & MAE $\downarrow$\\
      \midrule
      OCOO-T (short-panel, no patching) &
      \textbf{0.437} & 0.232 & 0.240 & 0.766 & \textbf{0.820} & \textbf{0.853} & \textbf{0.833} & \textbf{7.90e-3} & \textbf{0.061} \\
      OCOO-T (patch 8, vpred) &
      \textbf{0.437} & 0.244 & 0.239 & \textbf{0.772} & 0.816 & 0.834 & 0.754 & 1.06e-2 & 0.073 \\
      OCOO-T (patch 8, xpred) &
      0.435 & \textbf{0.248} & \textbf{0.242} & 0.769 & 0.811 & 0.831 & 0.744 & 1.08e-2 & 0.074 \\
      OCOO-T (patch 16, vpred) &
      0.401 & 0.235 & 0.233 & 0.744 & 0.803 & 0.819 & 0.744 & 1.03e-2 & 0.071 \\
      OCOO-T (patch 16, xpred) &
      0.428 & 0.244 & 0.239 & 0.766 & 0.816 & 0.834 & 0.756 & 1.06e-2 & 0.073 \\
      OCOO-T (patch 32, vpred) &
      0.434 & 0.243 & 0.239 & 0.767 & 0.802 & 0.819 & 0.706 & 1.09e-2 & 0.073 \\
      OCOO-T (patch 32, xpred) &
      0.411 & 0.240 & 0.230 & 0.759 & 0.815 & 0.832 & 0.800 & 1.04e-2 & 0.072 \\
      \bottomrule
    \end{tabular}
  }
\end{table}

\subsection{Cellular Context Conditioning: Covariate Embeddings vs. Mean Control-Cell Profiles}

Cellular context is a central conditioning signal for perturbation response prediction, because the same perturbation can induce substantially different transcriptional effects across cell lines, cell types, donors, and experimental batches. A simple and scalable strategy is to represent this context with learned covariate embeddings, such as cell-line or cell-type embeddings, and inject them into the denoising network together with perturbation and dosage conditions. This design is parameter-efficient and naturally supports in-silico generation once the relevant categorical covariates are known.

However, learned categorical embeddings are tied to the cellular contexts observed during training and therefore cannot directly represent new cell lines or cell types encountered at application time. On the Replogle-Nadig benchmark, we therefore compare this embedding-based conditioning strategy with a population-level alternative that uses the mean profile of a set of control cells as the cellular context condition, as described in Section 3.4. 

\begin{figure}[htb]
  \centering
  \includegraphics[width=0.92\linewidth]{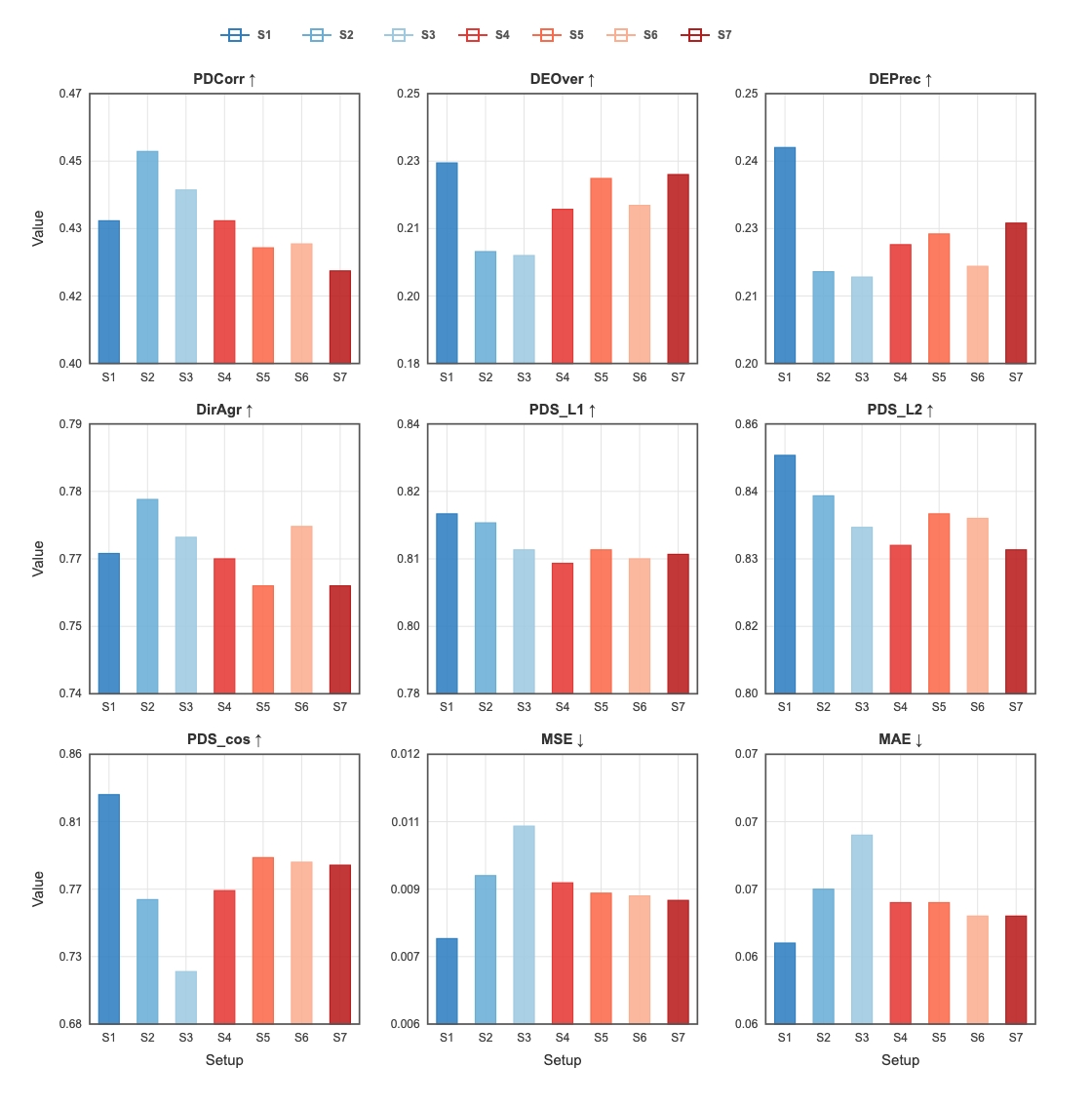}
  \caption{Comparisons of different cellular context injection methods on Raplogle-Nadig benchmark. S1: cell-line embeddings; S2--S7: mean control-cell profiles with set sizes 1, 4, 8, 16, 32, and 64, respectively.}
  \label{fig:vx_comp_bar}
\end{figure}

Figure~\ref{fig:vx_comp_bar} shows that using mean control-cell profiles as cellular context (S2-S7) can produce reasonably strong perturbation-response predictions across a range of control set sizes. This supports the motivation for population-level control-cell conditioning: even without a learned cell-line embedding, the averaged control profile carries substantial information about the basal cellular state and can guide the denoising model toward the appropriate perturbed distribution. Appendix C provides a more detailed analysis of MMD regularization during optimization, showing that adding MMD does not consistently yield further improvements. 

At the same time, the cell-line embedding setup (S1) achieves stronger performance on DE-related and PDS metrics, indicating that learned categorical context remains effective when the target cell line has been observed during training. Overall, these results suggest a trade-off between applicability and peak benchmark performance: mean control-cell profile conditioning offers a practical route for representing unseen or fine-grained cellular contexts, whereas learned cell-line embeddings can provide stronger perturbation-specific discrimination and differential-expression recovery in seen contexts.

\section{Discussion and Conclusion}

This work introduces \ocoot, a simple and scalable virtual cell model for transcriptional perturbation response prediction. Rather than relying on auxiliary autoencoders, specialized encoder-decoder architectures, or explicit distribution-matching modules, OCOO-T formulates perturbation response prediction as conditional continuous-time denoising directly in the original gene-expression space. At the architectural level, OCOO-T is built on a plain Transformer backbone: patched continuous expression profiles are processed by standard self-attention and feed-forward blocks, while perturbation identity, dosage, time, and cellular context are injected through lightweight conditioning mechanisms. Across chemical perturbation, genetic perturbation, and cytokine stimulation benchmarks, this minimalist Transformer design achieves strong performance on expression-level reconstruction, perturbation-discrimination, and differential-expression metrics.

The main empirical finding of this study is that architectural simplicity is sufficient to obtain competitive perturbation response prediction when the denoising objective and conditioning scheme are properly designed. Directly modeling continuous transcriptomic profiles avoids the information bottleneck imposed by pre-learned latent spaces and removes the need to train a separate cell-state encoder. The patching and depatching mechanism further makes long transcriptional profiles tractable for Transformer processing while preserving gene-level outputs. These results suggest that the proposed virtual cell model does not necessarily require increasingly complex auxiliary modules; instead, strong generative performance can emerge from a compact denoising backbone with biologically meaningful conditioning.

Although promising results have been observed, important gaps remain between current virtual cell benchmarks and real-world biological applications. In particular, although the Cell Eval toolkit provides a useful and increasingly standardized evaluation interface, existing metrics may still introduce biases and may not fully capture the biological utility of predicted perturbation responses in downstream scientific workflows. Future evaluation should therefore move beyond aggregate expression similarity and include criteria more directly tied to biological meaning, such as pathway-level consistency, regulatory-network recovery, and relevance to experimental decision-making. In addition, current benchmarks still provide limited coverage of out-of-distribution settings. More challenging OOD evaluations are needed to test whether perturbation models can generalize across unseen interventions, cellular contexts, laboratories, and experimental protocols. Addressing these challenges will be essential for turning virtual cell models from accurate benchmark predictors into reliable tools for biological discovery.

\clearpage
\appendix

\section*{Appendix}
\addcontentsline{toc}{section}{Appendix}

\section{Training Details}

Backbone of the denoiser model is a 12-layer Transformer with hidden size 768 and 12 attention heads (head dimension 64). Each block uses RMSNorm, QK-norm, SwiGLU feed-forward networks (expansion ratio 4.0), and adaLN-zero conditioning. Beginning at layer 4, 32 learnable in-context tokens carrying perturbation and cell-line conditioning are prepended to the sequence.

We train the Transformer denoiser network using AdamW with $\beta_1 = 0.9$, $\beta_2 = 0.95$, and no weight decay. The base learning rate is $5 \times 10^{-5}$ and is scaled linearly with the effective batch size according to
\begin{equation}
  \mathrm{lr} = \mathrm{blr} \times \frac{\mathrm{effective\ batch\ size}}{256}.
\end{equation}
We use a cosine decay schedule with a 5-epoch linear warmup and a minimum learning rate of 0. We maintain two exponential moving averages (EMAs) of the model weights with decay rates of 0.9999 and 0.9996; the second EMA is used for inference by default.

Each GPU processes a per-device batch size of 64 cells. With 8 NVIDIA H100 GPUs and no gradient accumulation, the effective batch size is 512, yielding an actual learning rate of $1 \times 10^{-4}$. Training is performed in BF16 mixed precision via torch.amp.autocast, and gradients are synchronized across GPUs using PyTorch Distributed Data Parallel (DDP).

Timesteps are sampled from a logit-normal distribution: $t = \text{sigmoid}(z)$ where $z \sim \mathcal{N}(P_\text{mean}, P_\text{std}^2)$ with $P_\text{mean} = -0.8$ and $P_\text{std} = 0.8$. This concentrates training on the noisier end of the trajectory. The model is trained under a rectified flow objective with velocity prediction (v-prediction), optimizing the L2 loss between the predicted and target velocity fields. During training, perturbation labels are randomly dropped with probability 0.1 to enable classifier-free guidance.

For sampling, we integrate the learned ODE using the Heun second-order solver over 50 uniformly spaced timesteps, with the final step replaced by an Euler step for numerical stability.

\section{Evaluation Metrics}
\label{app:evaluation_metrics}

We report both gene-expression-level metrics and differential-expression (DE) metrics to evaluate transcriptional response prediction quality. All metrics are computed via the Cell Eval toolkit released by STATE~\cite{arcinstitute2026celleval}.

\textbf{Gene-expression-level metrics.} These metrics are computed on pseudobulk expression profiles, obtained by averaging single-cell expression vectors under the same perturbation condition.

\begin{itemize}
  \item \textbf{PDCorr} (Pearson Delta Correlation) measures the agreement between the predicted perturbation effect and the true perturbation effect at the level of gene-wise expression changes. Let $\Delta_{\mathrm{real}} = X^{\mathrm{real}}_{\mathrm{pert}} - X^{\mathrm{real}}_{\mathrm{ctrl}}$ and $\Delta_{\mathrm{pred}} = X^{\mathrm{pred}}_{\mathrm{pert}} - X^{\mathrm{pred}}_{\mathrm{ctrl}}$. In our setting, the predicted control profile is taken to be the real control profile, so $X^{\mathrm{pred}}_{\mathrm{ctrl}} := X^{\mathrm{real}}_{\mathrm{ctrl}}$. PDCorr is then the Pearson correlation between $\Delta_{\mathrm{real}}$ and $\Delta_{\mathrm{pred}}$. Values closer to 1 indicate better agreement in perturbation-induced expression trends.
  \item \textbf{MSE} (Mean Squared Error) computes the mean squared difference between the predicted pseudobulk perturbed expression vector $X^{\mathrm{pred}}_{\mathrm{pert}}$ and the true vector $X^{\mathrm{real}}_{\mathrm{pert}}$. Lower values indicate more accurate absolute expression prediction.
  \item \textbf{MAE} (Mean Absolute Error) computes the mean absolute difference between $X^{\mathrm{pred}}_{\mathrm{pert}}$ and $X^{\mathrm{real}}_{\mathrm{pert}}$. Compared with MSE, MAE is less sensitive to outlier genes with extremely large expression values.
  \item \textbf{PDS\_L1}, \textbf{PDS\_L2}, and \textbf{PDS\_cos} (Perturbation Discrimination Scores) evaluate whether the predicted perturbation effect is most similar to the true effect of the same perturbation rather than to other perturbations. For each perturbation, the predicted effect vector is compared against the matrix of true effect vectors using L1 distance, L2 distance, or cosine distance, respectively. The distances are ranked, and the score is computed from the normalized rank of the correct perturbation match. Scores range from 0 to 1, with higher values indicating better perturbation specificity.
\end{itemize}

\textbf{Differential-expression metrics.} These metrics evaluate whether the model correctly identifies the genes that change significantly under perturbation.

\begin{itemize}
  \item \textbf{DEOver} (DE Overlap) measures the overlap between the predicted Top-$N$ DE genes and the true Top-$N$ DE genes, where ranking is based on absolute fold change under a significance threshold (default FDR 0.05). The score is the size of the intersection divided by $N$. Here N is valued by the number of real significant DE genes. Higher values indicate more accurate recovery of the most strongly changed genes.
  \item \textbf{DEPrec} (DE Precision) measures the precision of the predicted DE gene set. It uses the same overlap notion as DEOver, but the denominator is the number of predicted significant DE genes. This metric quantifies how many predicted significant genes are truly significant.
  \item \textbf{DirAgr} (Direction Agreement) evaluates whether the predicted direction of change for recovered DE genes matches the true direction of change. Specifically, it considers the intersection between true significant DE genes and predicted DE genes, compares the signs of their log fold changes, and reports the fraction with matching signs. Higher values mean the model more reliably predicts whether genes are up-regulated or down-regulated.
\end{itemize}

Together, these metrics assess complementary aspects of model quality: overall expression reconstruction, perturbation-specific discriminability, recovery of key DE genes, and correctness of the predicted direction of transcriptional change.

\section{Analysis of MMD-Regularized Flow Matching in High Dimensions}

We additionally explored augmenting the standard flow matching velocity prediction objective with a kernel-based Maximum Mean Discrepancy (MMD) regularizer. This design is motivated by recent single-cell perturbation studies showing that distribution-level matching can complement pointwise denoising supervision. In particular, scDFM ~\cite{yu2026scdfm} regularizes the terminal samples of a conditional flow matching model with a multi-kernel MMD loss, while PerturbDiff ~\cite{yuan2026perturbdiff} formulates an MMD-based denoising objective by embedding empirical cell populations into a reproducing kernel Hilbert space; both studies report that removing the MMD component weakens global distributional alignment and downstream differential-expression recovery. 

Specifically, in addition to the original flow matching objective, we introduce an MMD regularization term to encourage the generated perturbed-cell population to better match the empirical target distribution. For each training instance, we use the velocity network to compute a single-step estimate of the perturbed endpoint from the intermediate state:
\begin{equation}
\hat{x}_1
=
x_t + (1-t)\, v_{\theta}(z_t, p, c, b, t),
\end{equation}
This construction converts the learned velocity field into endpoint samples, enabling direct distribution-level regularization against the observed perturbed cells. We considered following auxiliary distributional loss of the form:
\begin{equation}
  \mathcal{L}
  =
  \mathcal{L}_{v}
  +
  \lambda\mathcal{L}_{MMD}
  =
  \mathcal{L}_{v}
  +
  \lambda \, \mathrm{MMD}^2(\hat{x}_1, x_1),
\end{equation}
where \(\hat{x}_1\) denotes the estimated terminal sample obtained from the predicted velocity, and \(x_1\) is the target sample. 

We evaluated several MMD-regularized variants by varying the kernel choices, including energy-based kernels and multi-scale Gaussian kernels, as well as the regularization weight \(\lambda\). We incorporate cellular context by sampling control cells directly from the control pool rather than using a precomputed population mean. This allows each minibatch to form an empirical approximation of the control population, aligning naturally with our minibatch-wise MMD regularization while preserving cell-to-cell variability. Empirically, MMD-regularized variants did not yield consistent improvement over the pure velocity-matching baseline. 
%In several settings, performance degraded. 
As shown in Table~\ref{tab:mmd-results}, 
%adding MMD did not yield a consistent improvement over the pure velocity-matching baseline. 
a small weight (\(\lambda=0.1\)) led to modest improvements in several DE and PDS related metrics, but reduced Pearson correlation and left MAE and MSE metrics essentially unchanged. Larger weights led to clear degradation across most evaluation criteria. 

\begin{table}[htb]
  \centering
  \caption{Replogle dataset evaluation results for different MMD loss setting.}
  \label{tab:mmd-results}
  \small
  \resizebox{\textwidth}{!}{%
    \begin{tabular}{lccccccccc}
      \toprule
      Model & \pdcorr $\uparrow$ & DEOver $\uparrow$ & DEPrec $\uparrow$ & DirAgr $\uparrow$  & PDS\_L1 $\uparrow$ & PDS\_L2 $\uparrow$ & PDS\_cos $\uparrow$ & MSE $\downarrow$ & MAE $\downarrow$\\
      \midrule
      Baseline & 0.443 & 0.234 & 0.232 & 0.767 & 0.780 & 0.800 & 0.728 & 0.007 & 0.058 \\
      % Baseline\_2 & 0.4445 & 0.2453 & 0.2477 & 0.7732 & 0.8217 & 0.8399 & 0.7392 & 0.0066 & 0.0574 \\
      MMD, $\lambda=0.1$ & 0.427 & 0.239 & 0.238 & 0.768 & 0.808 & 0.826 & 0.767 & 0.007 & 0.058\\
      MMD, $\lambda=1$ & 0.400 & 0.193 & 0.198 & 0.752 & 0.811 & 0.828 & 0.715 & 0.009 & 0.066\\
      MMD, $\lambda=5$ & 0.269 & 0.123 & 0.095 & 0.668 & 0.712 & 0.739 & 0.618 & 0.027 & 0.111\\
      \bottomrule
    \end{tabular}
  }
\end{table}

A plausible explanation is the known degradation of kernel-based discrepancy measures in high-dimensional regimes. Ramdas et al. ~\cite{ramdas2015decreasing} show under ``fair'' alternatives, where the KL divergence between distributions is kept constant as dimension increases, the test power or optimization signal of population MMD can decrease polynomially, and in some cases exponentially, with the data dimension. For instance, with an RBF kernel and median bandwidth, \(\mathrm{MMD}^2\) between Gaussian alternatives can scale as \(O(1/d)\). This suggests that MMD may provide a weak regularization signal in high-dimensional gene-expression spaces.

This observation is particularly relevant to our setting, where the model operates on gene-expression feature vectors with dimensionality around \(d=2000\). In such a regime, pairwise Euclidean distances tend to concentrate, and kernel values become dominated by global distance scale rather than informative local variation. Consequently, an MMD penalty applied directly in the raw gene-expression space may provide a weak or noisy gradient signal.

%Moreover, unlike the velocity loss, which provides sample-wise supervision for the conditional vector field, MMD is an unpaired minibatch-level distributional criterion. It encourages aggregate distributional matching between predicted and target samples, but does not directly preserve the conditional transport structure required by the flow matching objective.

Thus, the empirical degradation observed after adding MMD can be interpreted as a consequence of a mismatch between the auxiliary loss and the high-dimensional conditional generation problem. The MMD regularizer introduces a global distributional constraint whose effective signal may be attenuated in high dimensions, while also potentially conflicting with the pointwise velocity supervision. These results suggest that raw-space MMD is not necessarily an effective regularizer for high-dimensional flow-matching models. A more suitable use of kernel discrepancies may require applying MMD in a lower-dimensional or semantically structured representation space, such as PCA features, pathway-level features, or learned embeddings from a pre-train model.

%A more suitable use of kernel discrepancies may require applying MMD in a lower-dimensional or semantically structured representation space, such as PCA features, pathway-level summaries, learned embeddings from pre-train model, or selected differentially expressed genes. Alternatively, one may need carefully weighted or delayed MMD regularization, conditional MMD within perturbation groups, or kernels adapted to the geometry of gene-expression data. In our experiments, however, the pure velocity prediction loss provided a stronger and more stable training signal than MMD-augmented objectives.

\end{document}